\documentclass[runningheads]{llncs}
\usepackage{graphicx}
\usepackage{float}
\usepackage{mathtools}
\usepackage[bottom]{footmisc}
\usepackage[table,xcdraw]{xcolor}
\usepackage{booktabs}
\usepackage{multirow}
\usepackage{bbm}
\usepackage{subfig}
\usepackage{amsmath}

\usepackage{graphicx}
\usepackage[table,xcdraw]{xcolor}

\begin{document}
\title{ Coronavirus Detection and Analysis on Chest CT with Deep Learning
}
%
%
\author{Ophir Gozes\inst{1} \and
Maayan Frid-Adar \inst{1}\and
 Nimrod Sagie\inst{1} \and
 Huangqi Zhang\inst{2}\and \\ 
 Wenbin Ji\inst{2} \and
Hayit Greenspan\inst{1,3}}

\institute{
RADLogics Inc., Boston, MA\\ 
\and
Department of Radiology, Affiliated Taizhou Hospital of Wenzhou Medical University, , Zhejiang Province, China
\and
Department of Biomedical Engineering,
Tel-Aviv University
Tel-Aviv, Israel}

\authorrunning{Gozes et al.}
\titlerunning{}


\maketitle{}             
\begin{abstract}



The outbreak of the novel coronavirus, officially declared a global pandemic, has a severe impact on our daily lives.  As of this writing there are approximately 197,188 confirmed cases of which 80,881 are in “Mainland China” with 7,949 deaths, a mortality rate of 3.4\%. 
In order to support radiologists in this overwhelming challenge, we develop a deep learning based algorithm that can detect, localize and quantify severity of COVID-19 manifestation from chest CT scans.
The algorithm is comprised of a pipeline of image processing algorithms which includes lung segmentation, 2D slice classification and fine grain localization. In order to further understand the manifestations of the disease, we perform unsupervised clustering of abnormal slices. We present our results on a dataset comprised of 110 confirmed COVID-19 patients from Zhejiang province, China. 

\keywords{Corona \and Chest CT \and Lung \and COVID-19 \and AI \and Deep learning}

\end{abstract}

\section{Introduction}
The Coronavirus disease 2019 (COVID-19) is an infectious disease caused by severe acute respiratory syndrome coronavirus 2 (SARS-CoV-2). First identified in 2019 in Wuhan, China, it has since  become a global pandemic.
Although polymerase chain reaction (PCR) laboratory test is the gold-standard  for confirming COVID-19 positive patients, non-contrast thoracic CT scans have been shown as a potential tool in the disease detection \cite{mountsinai}. CT imaging studies of suspected population can support decision making, providing for immediate isolation and appropriate patient treatment.  The disease can be characterized by the presence of lung ground-glass opacities in early stages, followed by "crazy paving" and increasing consolidation. These findings led to the increase in CT scans in China, mainly in the Hubei province, eventually, becoming an efficient diagnosis tool.

It is well known that the ubiquitous Deep Learning based algorithms are data dependent. This is a challenge in scenarios in which a new category of events appears, and data is scarce. Moreover, since knowledge of disease characteristics in the early stage of spread of disease is limited,  the ability to perform exact annotation of the data to the various radiological manifestation is limited. 
In this work, we show the methods researched in the field of medical image computing - particularly image clustering, segmentation, and classification  can be harnessed to accurately and more rapidly assess disease progression and guide therapy and patient management.
In order to quickly come up with solutions, algorithms which can generalize from relatively low number of samples with weak supervision are required. For this reason, we base our system architecture on a sequence of 2D processing followed by 3D fusion stage.
One advantage of 2D processing is that it allows leveraging of existing pre-trained architectures thus reducing the time and resources required to train a deep learning model\cite{MetaChexnet}. 

\begin{figure}[t]
\centering
\includegraphics[width=12.0 cm]{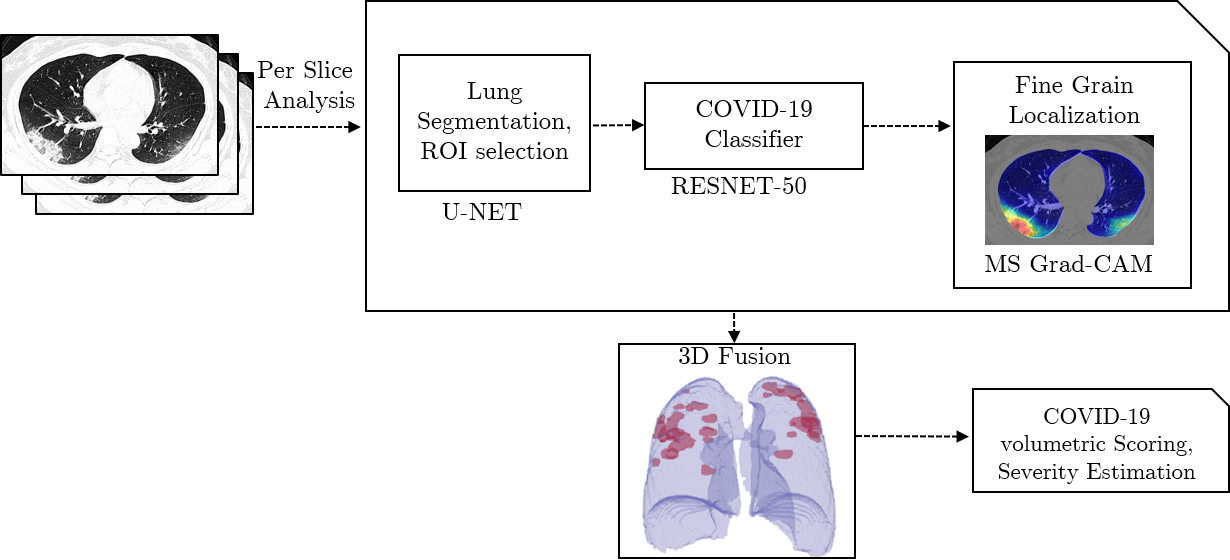}
\caption{Block diagram of the system}
\label{fig:blocks}\end{figure}

The solution we propose receives non-contrast chest CT scans and detects cases suspected with COVID-19 features. For cases classified as positive, the system outputs a lung abnormality localization map and a score related to the disease severity. Figure \ref{fig:blocks} shows a block diagram of the developed system, which we will describe in detail in the following sections.

\section{Methods}

 The proposed system is comprised of a number of processing steps:  We start with localizing the lung region of interest in the Chest CT. The second step utilizes  a 2D ROI classification network to classify the lungs ROIs as normal vs abnormal (COVID-19).
Using a multi-scale application of the GradCam \cite{gradcam} method,  a fine-grained map of the pathological tissue is extracted. The concatenation of  slice detection results as well as the localization maps to a single volume,  enables the creation of a Corona score which is a volumetric measure of the disease extent.
The advantage of using activation maps for estimating the disease severity grade lies in the fact that it is a data driven approach to localize disease manifestations. Thus it requires only weak slice based annotations.
Finally, in order to learn the different patterns of the abnormal manifestation of the disease we propose unsupervised clustering of the normal and abnormal slices. 
Several datasets  are used throughout, for development and testing, as specified in Table 1. 



%

\begin{table}[]

\resizebox{\textwidth}{!}{%
\begin{tabular}{@{}l|l|l@{}}
\toprule

Nº &
  Dataset &
  Description \\ \midrule
1 &
  \begin{tabular}[c]{@{}l@{}}Development Dataset\\ Source: Chainz \cite{chainz} \end{tabular} &
  \begin{tabular}[c]{@{}l@{}}50 abnormal thoracic CT scans (slice thickness, \{5,7,8,9,10\} mm) from China of
  patients \\ that were  diagnosed by a radiologist as suspicious for COVID-19 (from Jan-Feb 2020). \\ The cases were extracted by querying a cloud PACS system for cases that were referred\\  for laboratory testing following the scan. \\ Cases were annotated for each slice as normal (n=1036) vs abnormal (n=829)\end{tabular} \\ \midrule
2 &
  \begin{tabular}[c]{@{}l@{}}Testing Dataset\\ Source: Zhejiang Province, China\end{tabular} &
  \begin{tabular}[c]{@{}l@{}} 109 patients with confirmed diagnosis of COVID-19 infection by RT-PCR \\ 90 Patients with fever symptoms and upper respiratory tract symptoms diagnosed by \\ radiologist as Negative for COVID-19

  \end{tabular} \\ \midrule

3 &
  \begin{tabular}[c]{@{}l@{}}Lung segmentation Development\\ Sources: El-Camino Hospital (CA)\end{tabular} &
  6,150 CT slices of cases with lung abnormalities and their corresponding lung masks \\ \midrule
4 &
  \begin{tabular}[c]{@{}l@{}}Lung segmentation Development\\University Hospitals of Geneva (HUG).\end{tabular} &
  \begin{tabular}[c]{@{}l@{}}ILD database - A multimedia collection of cases with interstitial lung diseases (ILDs)\\  built at the University Hospitals of Geneva (HUG).\\  The dataset contains high-resolution computed tomography (HRCT) image series\\  with three-dimensional annotated regions of pathological lung tissue along \\ with clinical parameters from patients with pathologically proven diagnoses of ILDs.\end{tabular} \\ \bottomrule
\end{tabular}%

}

\caption{Datasets used in this work}
\end{table}

\subsection{Lung Segmentation and lung ROI selection}

\begin{figure}
\centering
\includegraphics[height=5.0 cm]{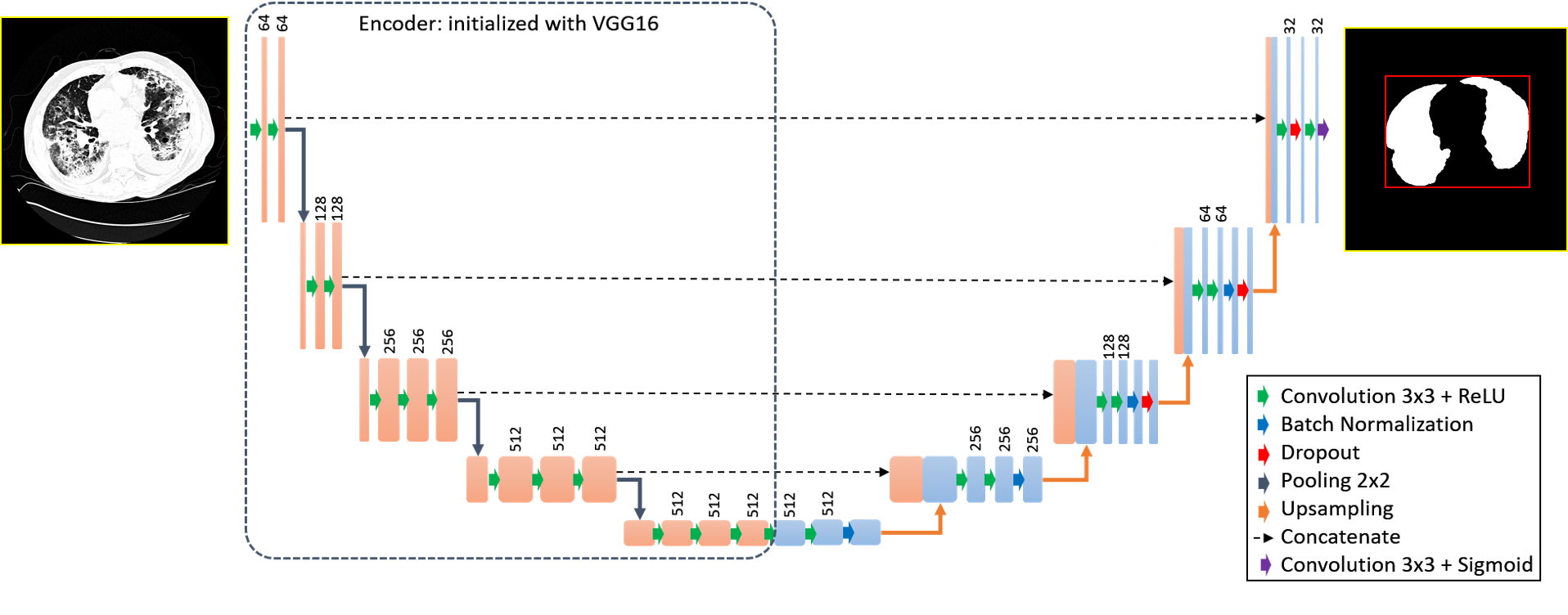}
\caption{The proposed U-Net architecture with a VGG-16 based encoder.}
\label{fig:architecture}\end{figure}

The purpose of this stage is to provide 2D lung segmentation to support visualization and ROI selection as input for the classification stage.  The model is based on the U-net architecture \cite{pretrained_unet}\cite{unet_olaf} in which the encoder is  pre-trained with ImageNet. Since the algorithm is required to support cases of abnormal lung appearance, we train it on a collection of slices belonging to datasets containing interstitial lung disease cases and opacities.
We use 6,150 CT slices of cases with lung abnormalities and their corresponding lung masks which were taken from a U.S based hospital (Table  1: Dataset-3,4). Input slices are normalize to a windows of [-1000, 0] HU. A diagram of the architecture is displayed in Fig. \ref{fig:architecture}. 
For ROI selection, we predict a segmentation mask for each slice of the case and take the largest bounding box to create the lung crop. This allows the classifier to focus the learning process on lung related areas.

\subsection{COVID-19 Classifier}

The second step focuses on classifying the lung ROIs as normal vs abnormal (COVID-19). 
We use a ResNet-50 - 2D deep convolutional neural network architecture \cite{resnet50} pretrained on ImageNet \cite{Imagenet}. The input size is 224$\times$224.
To train the network, COVID-19 cases suspected for COVID-19 from several Chinese hospitals are used (Table. 1: Dataset-1). From these cases,  1865 slices were annotated per as normal (n=1036) vs abnormal (n=829) for Coronavirus. We randomly split the dataset to 1725,320,270 slices for training, validation, and testing respectfully with no patient overlap between the test set and the development sets. To overcome the limited amount of cases, we employ data augmentation techniques (image rotations, horizontal flips and cropping). The binary cross entropy loss function was optimized using Adam Optimizer(lr=1E-4). 

\subsection{Model Interpretation and multi-scale GradCam for Localization }
\begin{figure}
\centering
\includegraphics[height=5.0 cm]{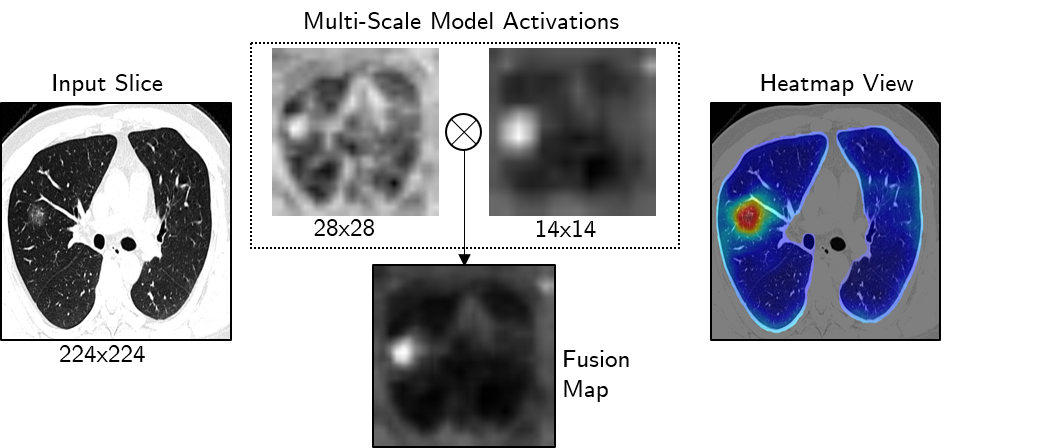}
\caption{Creation of fine-grain localization maps. The resulting heatmap view is a fusion of activation maps of two scales}
\label{fig:gradcam}\end{figure}

The GradCam  method has become a common tool for providing visual explanations when using Deep Networks with Gradient-based localization. In order to verify that the learning process focused on pathological areas, we employ the GradCam technique in two resolutions. GradCam is performed on the activation layers corresponding to resolution of 14X14 and 28x28 following a normalization to range [0,1]. The two maps are combined by multiplication to achieve a fine grain localization map.
In Fig.\ref{fig:gradcam} we display the activation maps corresponding to the two scales and the resulting heatmap obtained by fusion.

\subsection{Case-wise COVID-19 Scoring for Severity Estimation  }

To provide a complete review of a case, we combine the fine grain maps to create  3D localization maps. The resulting volume corresponds to the extent of the disease spread throughout the lungs and can provide valuable insight to the radiologist (Fig.\ref{fig:blocks}). In order to extract a quantitative metric we propose the \emph{\textbf{corona score}} (Eq.\ref{eq:corona_score}). The score is computed by summation over the  activation maps ($Cmap_{z}$) of positive detected slices, considering only activation above a certain pre-defined threshold $T_{activation}$. The threshold used in our experiments was determined to be 0.6 by means of visual evaluation by expert.

\begin{equation}
     \sum_{z=1} \sum_{ij} Cmap_{zij}\cdot \mathbbm{1}(cmap_{zij}>T_{activation})\cdot voxel\_volume
     \label{eq:corona_score}
\end{equation}

\section{Experiments and Results}
\label{exp_results}

\subsection{Slice classification results and experiments}
We start with an evaluation of the ability to detect slice-level Coronavirus. The performance of this step is crucial for obtaining overall case wise detection. For the validation step, we used 10\% of the slices from the development dataset comprised of cases from the Chinese population (Table I: Dataset-1). The split was patient wise and there is no overlap with the slices used for training. A total of 270 slices were analyzed: 150 normal slices and 120 COVID-19 suspected slices. We achieved an Area Under Curve (AUC) result of  0.994 with 94\% sensitivity and 98\% specificity (at threshold 0.5).

\subsection{Corona score experiments}
\subsubsection{Corona score for disease detection}
We use the corona score to classify Coronavirus vs non-Corona virus patients at the case-level.  
We use a collection of 109 patients with confirmed COVID-19 diagnosis and 90 non-Coronavirus patients (Table I: Dataset-2) from Zheijang, China. 
We performed an ROC analysis using the Corona score as a predictor variable. Results are shown in Fig.\ref{Fig:corona_graphs}.a :	
We achieve an AUC score of 0.948 (95\%CI: 0.912-0.985) on Chinese control and infected patients.

\subsubsection{Corona score as a criteria for severity }
\begin{figure}[h]
  \centering
    \subfloat[]{\includegraphics[width=0.55\textwidth]{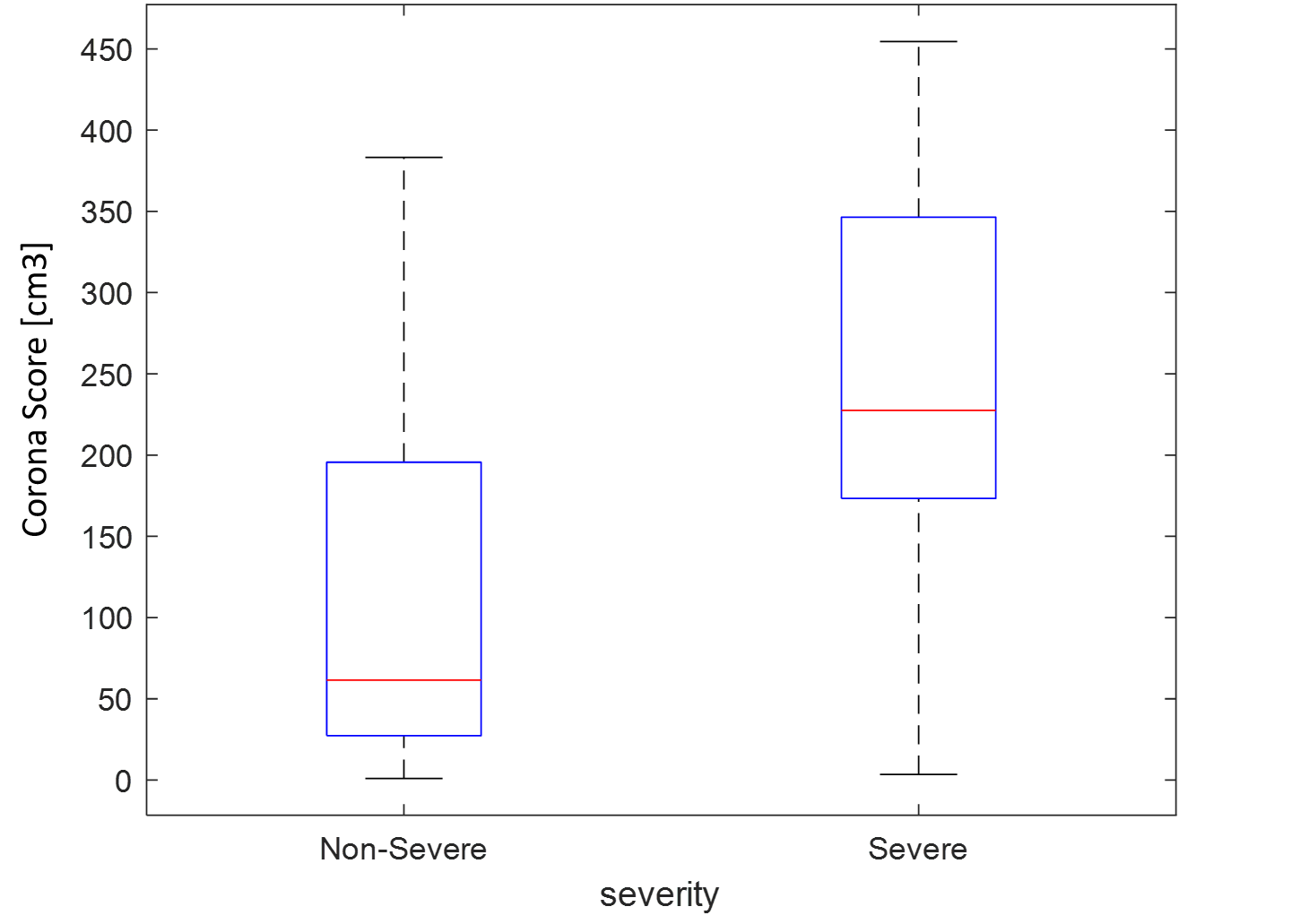}\label{Fig:corona_score_graph}}
  \hfill
  \subfloat[]{\includegraphics[width=0.4\textwidth]{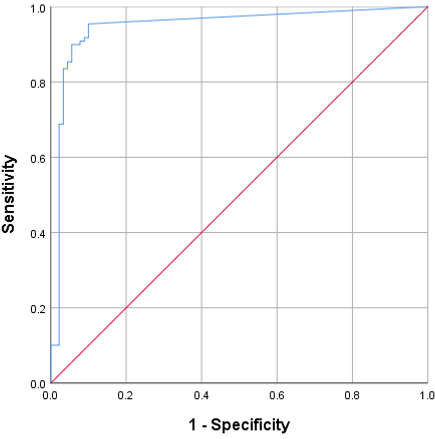}\label{Fig:ROC_analysis}}
  \caption{(a) Corona Score volumetric measure for non-severe and severe-grade Coronavirus patients;
  (b) ROC analysis for COVID-19 patient classification.}
  \label{Fig:corona_graphs}
\end{figure}

The corona score is a volumetric score related to the extent of the disease in a patient's lungs.
In order to examine the corrspondence between our metric and the radiologist's metric, we performed the following experiment:
A collection of 49 patients belonging to Dataset 2 were classified by a radiologist to  severe (n=13) vs non-severe (n=36). 
For the analysis, we take the corona score from the first time point of each patient and create a box-plot diagram comparing the score distributions of severe and non-severe patients. Results are shown in Fig.\ref{Fig:corona_graphs}.b :  The median corona score for the severe and non-severe cases was 61.5 cm3, and  227.5 cm3, respectively. The p-value of a two-sided Wilcoxon rank sum test for equality of distributions was (p=0.0064).



\subsection{Analysis of the disease manifestations}
In an initial study to explore and  identify prominent manifestations of the Coronavirus disease we use unsupervised k-means feature clustering. The feature space created by the COVID-19 classifier was selected as it encodes information relevant to the pathology and is more specific then ImageNet pre-trained feature extractors.
To further focus on disease related features, we weight each feature vector by the gradient of this layer computed using GradCam.
Prior to clustering, we perform normalization to zero mean and unit standard deviation. 

K-means clustering was performed where the optimal number of clusters (k=3) was found using the elbow method. 
For each input slice we extract the features by flattening the output of the last convolutional layer to a d=2048 feature vector, followed by element-wise multiplication with the corresponding gradients vector and pre-clustering normalization. 

Our analysis was conducted on positive key slices from 110 patients (n=1592) and on negative slices from 81 patients (n=701). For visualization we perform dimensionality reduction using Principal Components Analysis (PCA) from d=2048 to d=2. 

\begin{figure}[h]
\centering
\includegraphics[height=10cm,trim=4cm 0cm 0cm 0cm,clip]{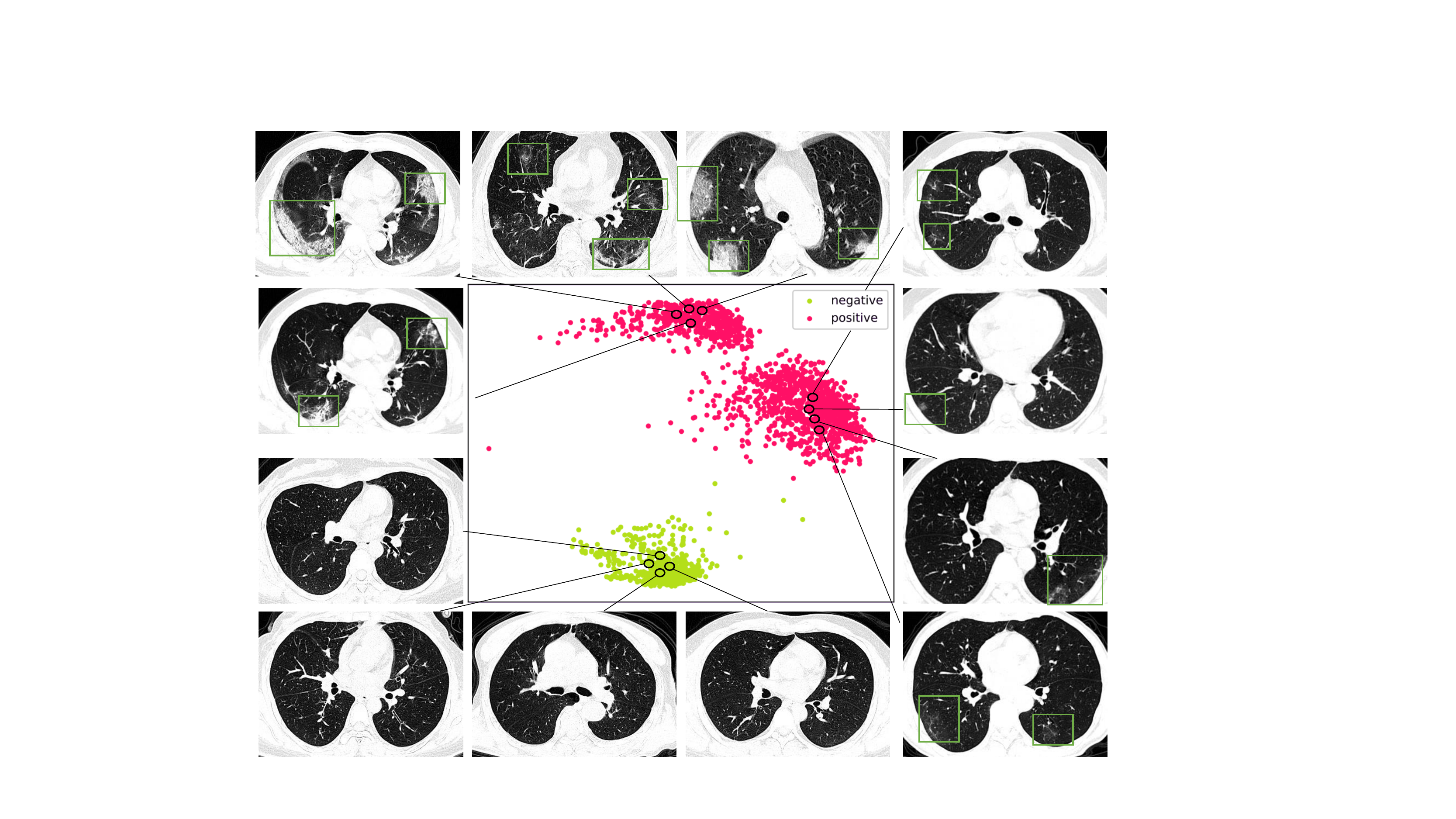}
\caption{Gradiaent weighted Feature Space and representative key slices}
\label{fig:corona_space}\end{figure}

 The resulting feature space is visualized in Fig. \ref{fig:corona_space}:  each point represents a slice in the feature space belonging to Negative and Covid-19 Positive slices. For each cluster we present 4 key slices chosen by minimal L2 distance to the corresponding cluster's center. It is noticeable that the negative samples group is well distinguished from the positives. Furthermore, the positive data points can be divided into two distinguishable groups: slices with subtle focal lesions and slices with diffuse severe patterns.

\section{Discussion}

In this study, we show results for Coronavirus detection, quantification and exploration using weakly-supervised deep-learning techniques on Chest CTs. 
This is the one of the first reports to our knowledge in this domain.

Utilizing the deep-learning image analysis system developed, we achieved classification results for Coronavirus vs Non-coronavirus cases per thoracic CT studies of 0.948 AUC (95\%CI: 0.912-0.985) on a dataset of Chinese control and infected patients. The fine-grain localization can be attractive to the radiologists by offering insight into the “black-box” model prediction as well as visualization on the 3D case-level.

The conducted exploratory analysis of the learned COVID-19 feature space, has identified main clusters in which slices reside. These clusters correspond to different disease manifestations (Normal, Focal, Diffuse). The value of this  approach is in the automatic exploration of a new unknown disease by means of an unsupervised algorithm.

Finally, we demonstrated that the suggested “Corona score” corresponds to the clinical grade of the disease; We therefore hypothesize that such a score may provide a means for patient monitoring and management.  

\end{document}